\documentclass[pre,superscriptaddress,twocolumn]{revtex4}
\usepackage{epsfig}
\usepackage{latexsym}
\usepackage{amsmath}
\begin {document}
\title {Synchronization in coupled map lattices as an interface depinning}
\author{Adam Lipowski}
\affiliation{Department of Physics, University of Geneva, CH 1211
Geneva 4, Switzerland}
\affiliation{Faculty of Physics, A.~Mickiewicz University,
61-614 Pozna\'{n}, Poland}
\author{Michel Droz}
\affiliation{Department of Physics, University of Geneva, CH 1211
Geneva 4, Switzerland}
\begin {abstract}
We study an SOS model whose dynamics is inspired by recent studies of the 
synchronization transition in coupled map lattices (CML).
The synchronization of CML is thus related with a depinning of interface from a 
binding wall.
Critical behaviour of our SOS model depends on a specific form of binding
(i.e., transition rates of the dynamics).
For an exponentially decaying binding the depinning belongs to 
the directed percolation universality class.
Other types of depinning,  including the one with a line of 
critical points, are observed for a power-law binding.
\end{abstract}
\maketitle
Recently synchronization of chaotic systems attracted a lot of
interest~\cite{FUJISAKA}.
To a large extent, this interest is motivated by numerous experimental 
realizations of this phenomena including lasers, electronic circuits, or
chemical reactions~\cite{BOCCAL}.
Synchronization acquires additional features in spatially extended systems, 
where it can be regarded as a certain nonequilibrium phase transition.
There are increasing efforts to understand the properties of this 
transition.
Relatively well understood is the synchronization  transition (ST) in certain
cellular automata.
Since a synchronized state can be regarded as an absorbing state, for cellular
automata, which are discrete systems, the phase transition, as expected, 
belongs to the Directed Percolation (DP) universality class~\cite{GRASS}.
However, continuous systems, as e.g., coupled map lattices (CML)~\cite{KANEKO}, need 
infinite time
to reach a synchronized state and the relation with DP does not seem to hold.
Indeed, Pikovsky and Kurths argued~\cite{PIKKURTHS} that 
for continuous systems this transition should belong to the so-called 
bounded Kardar-Parisi-Zheng (BKPZ) universality class~\cite{TU}.
Recently, precise numerical calculations confirmed their predictions but only
for CML with some continuous maps~\cite{AHLERS}.
Surprisingly, the ST for discontinuous~\cite{LIVI,AHLERS}, or continuous but 
sufficiently steep~\cite{LIPDROZ} maps was found to belong to the DP 
universality class.
It would be desirable to understand the critical behaviour of the ST in CML 
and the present paper might be a step in this direction.

Let us briefly describe the setup which is used to study
synchronization in CML~\cite{AHLERS}.
In the simplest, one-dimensional case, one takes a single-chain CML of size 
$L$ that is composed of $L$ diffusively coupled local maps 
$f(u_i)$~\cite{KANEKO}.
The maps are chaotic and act on continuous site variables 
$u_i$ ($i=1,\ldots,L$), that are typically bounded ($0<u_i<1$).
Then, one couples such a spatio-temporally chaotic system with its 
identical copy, which initially has a different set of site variables.
It turns out that the evolution of such a coupled system depends on the 
coupling strength.
For weak coupling the two CML's are desynchronized and essentially independent.
However, for sufficiently strong coupling the system gets synchronized and 
approaches a state where corresponding pairs of site variables 
in both copies take the same values.
To quantify synchronization one can introduce the  synchronization error 
$w(i,t)=|u_1(i,t)-u_2(i,t)|$ where a lower index denotes a copy of CML, and
its spatial average $w(t)=\frac{1}{L}\sum_{i=1}^L w(i,t)$.
To relate this problem with BKPZ one argues that in the continuous limit and 
close to the synchronized state, the
evolution equation for $w(i,t)$ is given as a Langevin equation with 
multiplicative
Gaussian noise which then, using the Hopf-Cole transformation $h=-{\rm ln}(w)$, is
transformed into BKPZ.
Let us notice that CML's are deterministic systems and the noise has only 
effective meaning, mimicking their chaotic behaviour.
The problem of a relation of such deterministic systems with stochastic 
counterparts is very interesting and recently is drawing some 
attention~\cite{EGOLF}.
In the above representation the desynchronized phase in CML ($w(t)>0$) 
corresponds to 
the interface pinned relatively close to the wall ($\langle h_i\rangle <\infty$).
In the synchronized phase ($w(t)\rightarrow 0$) the interface depins and 
drifts away ($\langle h_i\rangle\rightarrow \infty$).
Perfectly synchronized state ($\langle w(t)\rangle=0$) is reached only 
after infinitely 
long time.
The above analysis requires the differentiability of the local map and thus is 
not applicable to discontinuous maps.
However, numerical results show that the relation with BKPZ breaks down also 
for continuous but sufficiently steep maps~\cite{LIPDROZ}.
From a theoretical point of view it would be desirable to understand why such
properties of the local map affect the nature of the ST and move into the DP
universality class.
Let us notice, that if the coupled CML system enters the synchronized 
state, it will remain in this state forever.
Such a state can be thus considered as an absorbing state of the dynamics, 
although it cannot be reached in any finite time.
Well-developed techniques are available to study phase transitions 
in models with absorbing states~\cite{HAYE}.

It is clear that the problem of synchronization in extended systems is related 
with a number of very interesting problems in nonequilibrium statistical 
mechanics such as the KPZ model, nonequilibrium wetting~\cite{PAST} or 
directed percolation.
Recently, some arguments were given that a notoriously difficult particle system,
the so-called PCPD model, might be also related with these models\cite{HIN}.
Deeper understanding of these problems and their mutual relations would be 
certainly desirable.

In the present paper we introduce an interfacial, discrete (SOS) model which
is inspired by the dynamics of CML with discontinuous maps.
The synchronization of CML is thus related with a depinning of interface from a 
binding wall.
Numerical calculations show that the universality class of the depinning 
transition in our model depends on the choice of binding, which enters the 
dynamics through certain transition rates.
For an exponentially decaying binding, the depinning belongs to the DP 
universality class.
In this case  the overall behaviour of the model is very similar to 
Sneppen's model of interface propagation 
in a random environment that is driven by a certain extremal 
dynamics~\cite{SNEPPEN,OLAMI}.
For a power-law decaying binding the depinning transition is characterized by a 
different set of critical exponents.
In the case of a rapid decay these exponents are very close to those obtained for the 
bosonic version of PCPD model~\cite{KOCKEL}.
In the case of a slow decay, in the entire unbounded phase the interface remains 
critical, that in itself is an interesting property of a nonequilibrium system.

In our model discrete site variables $h_i=1,2,\ldots$  are defined on a 
one-dimensional lattice of size $L$ ($i=1,\ldots,L$) with periodic boundary
conditions ($h_{L+1}=h_1$).
In an elementary update we select randomly a site $i$ and change the variable
$h_i$, and possibly neighbouring ones, according to the following rule: 
(i) with probability $p(h_i)$ one sets $h_i=h_{i+1}=h_{i-1}=1$; (ii) with 
probability $1-p(h_i)$ the site variable $h_i$ increases by unity 
($h_i\rightarrow h_i+1$).
During a unit of time $L$ elementary updates are performed.
To complete the definition we have to specify the function $p(h)$.
To allow a drift toward $h=\infty$ the function $p(h)$ must decay to zero for 
$h\rightarrow\infty$.
Numerical results that we present below are obtained for two cases: 
$p(h)={\rm e}^{-\gamma h}$ (model I) and $p(h)=a(h+1)^{-\gamma}$ 
(model II), where $\gamma>0$ and $a>0$ are control parameters of the model.

To make a link with synchronization in CML's the following remarks are in 
order.
Numerical calculations for CML's with discontinuous maps show that the
Lyapunov exponent
that governs the evolution of the synchronization error $w(i,t)$ is 
negative in the vicinity of the transition~\cite{AHLERS}. 
Approximately, the evolution of $w(i,t)$ is thus made of 
consecutive contractions ($w(i,t)\rightarrow cw(i,t)$ and $c<1$) that, due to 
discontinuity of the map, are 
from time to time interrupted by discontinuous changes that might substantially
increase the value of $w(i,t)$.  
To notice a link between the dynamics of synchronization error in CML and
our model we introduce new variables $w_i={\rm e}^{-h_i}$ 
(let us notice a similarity to the inverse Hopf-Cole transformation).
Indeed, the increase of $h_i$ by unity according to the rule (ii) decreases 
$w_i$ by a factor e that corresponds to the 
contraction of $w(i,t)$.
The first rule mimics the discontinuous jumps of $w(i,t)$. 
Since local maps in CML's are coupled, a jump at a site $i$ also affects its 
neighbours.

Monte Carlo simulations of our model are similar to those of other models with 
absorbing states~\cite{HAYE}.
For some details related to the fact that the model needs an infinite time
to reach an absorbing state see e.g.,~\cite{LIPDROZ}.
We observed that for sufficiently large $\gamma$ the interface depins from the 
$h=1$ wall and drifts away.
For smaller $\gamma$ the model remains in the active phase with the interface 
relatively close to the wall.
To examine the nature of the phase transition we introduced 
$w(t)=\langle \frac{1}{L}\sum_{i=1}^L w(i,t)\rangle$ that in the steady state
is denoted as $w$ (in the following we refer to this quantity as activity).
Of course in the absorbing phase $w=0$ and in the active phase $w>0$.
Upon approaching the critical point $w$ typically exhibits a power-law decay
$w\sim(\gamma-\gamma_c)^{-\beta}$, with a characteristic exponent $\beta$ and
for the critical point located at $\gamma=\gamma_c$.
Moreover, we studied the time dependence of $w(t)$.
One expects that at criticality this quantity has a power law
decay $w(t)\sim t^{-\Theta}$, where $\Theta$ is another characteristic exponent.
We also used the so-called spreading technique~\cite{SPREADING}.
First, we set $h_i=\infty$ for every but one site ($i_0$) that was set to 
unity.
Then we monitored the subsequent evolution of the model (actually, of interest 
are only sites with positive $w(i,t)$ i.e., with finite $h_i$) measuring
the average activity of the system  $w(t)$, the survival probability $P(t)$,
and the averaged spread square $R^2(t)=\frac{1}{w(t)}\sum_i w(i,t)(i-i_0)^2$.
One  expects that at
criticality:
$w(t)\sim t^{\eta}$, $P(t)\sim t^{-\delta}$, and $R^2(t)\sim t^{z}$, that 
at the same time defines the critical exponents $\eta$, $\delta$, and $z$.
\begin{figure}
\centerline{\epsfxsize=8cm
\epsfbox{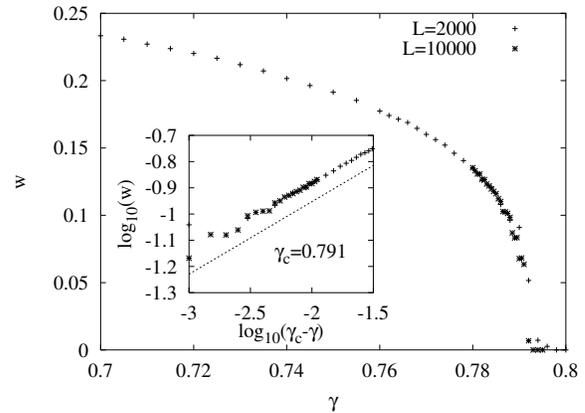}
}
\caption{
The steady state activity $w$ as a function of $\gamma$ for model I.
The inset shows the logarithmic scaling of $w$ at criticality.
The slope of the dotted line corresponds to the DP value $\beta=0.2765$.
The simulation time was $t_{{\rm sim}}=10^5$ plus $2\cdot 10^4$ discarded for
relaxation.
}
\label{roe}
\end{figure}
\begin{figure}
\vspace{-1.5cm}
\centerline{\epsfxsize=8cm
\epsfbox{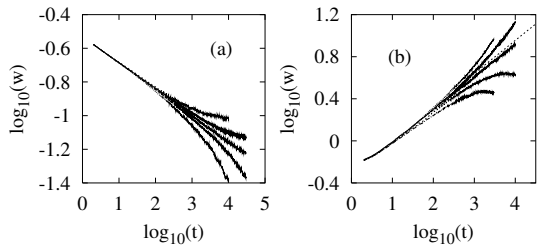}
}
\caption{ (a) The time-dependent activity $w(t)$ as a function of time $t$ for
(from top) $\gamma=0.788$, 0.790, 0.791 (critical point), 0.792, and 0.794
(Model I).
The results are averaged over 100 independent runs and simulations were done 
for $L=5\cdot 10^4$.
(b) The time-dependent activity $w(t)$ as a function of time $t$ calculated
using the spreading method for
(from top) $\gamma=0.787$, 0.789, 0.791 (critical point), 0.793, and 0.795
(Model I).
The results are averaged over $10^5$ independent runs.
The slope of the dotted line corresponds to the DP value $\eta=0.3137$
}
\label{timee}
\end{figure}

First, let us describe the results obtained for model I 
($p(h)={\rm e}^{-\gamma h}$).
From the steady state measurements of $w$ (Fig.~\ref{roe}) we estimate 
$\gamma_c=0.791(1)$, and $\beta=0.28(1)$.
Such a location of the critical point is confirmed from time-dependent 
simulations (Fig.~\ref{timee}a) and the spreading method (Fig.~\ref{timee}b).
From these data we estimate $\Theta=0.160(2)$, $\eta=0.317(5)$, 
$\delta=0.16(1)$, and $z=1.26(1)$.
The results for $P(t)$ and $R^2(t)$ are not presented.
Obtained values of critical exponents  clearly show that model I belongs to
the DP universality class for which~\cite{HAYE}: $\beta=0.2765$, $\eta=0.3137$,
$\Theta=\delta=0.1595$ and $z=1.265$.

\begin{figure}
\centerline{\epsfxsize=8cm
\epsfbox{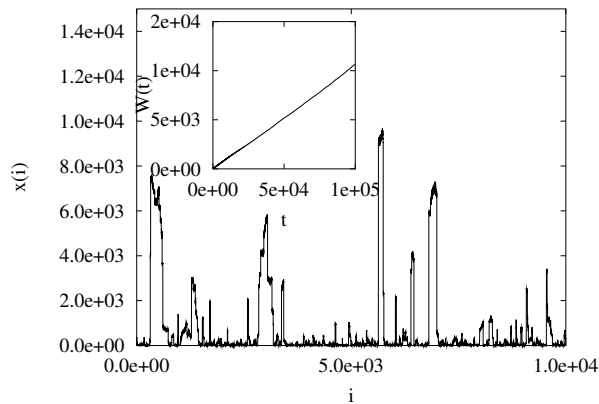}
}
\caption{
The interface profile for model I at criticality ($\gamma=\gamma_c=0.791$) 
after $t=10^4$ MC steps.
The inset shows that the interface width $W(t)$ grows linearly in time.
}
\label{conf}
\end{figure}

To have a more complete insight into the critical behaviour of our model we
calculate the average interfacial width 
$W(t)=\langle [\sum_{i=1}^L (h(i,t)-\langle h(i,t)\rangle)]^2\rangle^{1/2}$.
At criticality this quantity typically behaves as $W(t)\sim t^{\beta'}$ where
$\beta'$ is the growth exponent.
For model I simulations show almost linear increase of $W(t)$ with time
(inset in Fig.~\ref{conf}) and we estimate $\beta'=1.0(1)$.
Such a value of the growth exponent indicates very strong fluctuations, 
which is confirmed through a visual inspection of the interface profile 
(Fig.~\ref{conf}).
In principle, from direct calculations for CML's we can obtain an interface 
profile as a logarithm of the synchronization error.
Our preliminary calculations for CML model examined by Ahlers and 
Pikovsky~\cite{AHLERS} show that large fluctuations, similar to those 
in model I are clearly seen.
However, to quantify these fluctuations and calculate e.g., 
the width $W(t)$, simulations on a longer time scale are required.
Since for CML dynamics is defined in terms of $w$'s rather than $h$'s,
simulations for longer time, which must probe states of extremely small 
synchronization error, severely suffer from the finite accuracy of 
numerical computations.
On the other hand our SOS model easily allows us to examine such a 
long-time regime.

It is already known that some SOS-like models belong to the DP universality 
class.
In particular,  Alon et al.~\cite{DPWETT} introduced a model of interface 
roughening with edge evaporation.
In this model the growth of the interface width is logarithmic in 
time and is thus much different than in our model.
As far as the critical behaviour is concerned, 
our model seems to be more closely related with the Sneppen's model
of interface spreading in environment with quenched disorder~\cite{SNEPPEN}.
His model is driven by extremal dynamics and has the growth exponent 
$\beta'$ close to unity.
Later it was established that this model is actually equivalent to the DP 
model sitting exactly at the critical point~\cite{OLAMI}.
Important ingredients of Sneppen's model are quenched 
disorder and extremal dynamics.
The fact that our model, that misses these features, exhibits essentially the
same behaviour is in our opinion quite interesting and worth further 
examination.
\begin{figure}
\centerline{\epsfxsize=8cm
\epsfbox{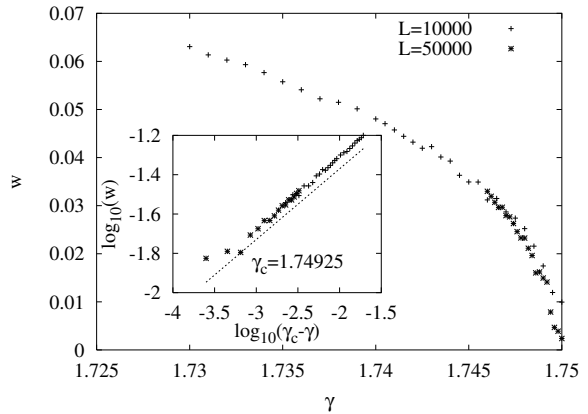}
}
\caption{
The steady state activity $w$ as a function of $\gamma$ for model II for $a=1$.
The inset shows the logarithmic scaling of $w$ at criticality.
The slope of the dotted line corresponds to $\beta=0.36$.
The simulation time was $t_{{\rm sim}}=10^5$ plus $2\cdot 10^4$ discarded for
relaxation.
}
\label{ro}
\end{figure}
\begin{figure}
\vspace{-1.5cm}
\centerline{\epsfxsize=8cm
\epsfbox{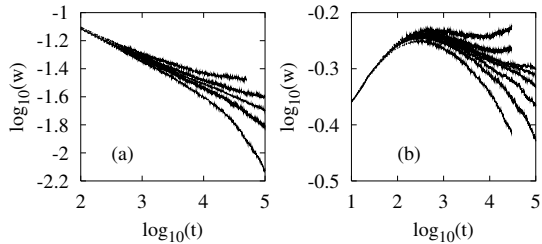}
}
\caption{
(a) The time-dependent activity $w(t)$ as a function of time $t$ for
(from top) $\gamma=1.746$, 1.748, 1.74925 (critical point), 1.750, and 1.752
(Model II, $a=1$).
The results are averaged over 100 independent runs and simulations were done 
for $L=5\cdot 10^4$.
(b) The time-dependent activity $w(t)$ as a function of time $t$ calculated
using the spreading method for
(from top) $\gamma=1.7488$, 1.749, 1.7492, 1.74925 (critical point), 1.7493, 
1.7494, 1.7496 and 1.75 (Model II, $a=1$).
The results are averaged over $10^5$ independent runs.
}
\label{time}
\end{figure}

Using the same procedure we studied the model II ($P(h)=a(h+1)^{-\gamma}$).
First we kept $a=1$ fixed and varied only the parameter $\gamma$.
Some of our numerical results are shown in Fig.~\ref{ro}-Fig.~\ref{time}.
Using these data we estimate: $\gamma_c=1.74925(5)$, $\beta=0.36(1)$, 
$\Theta=0.185(3)$, $\eta=-0.03(1)$, $\delta=0.445(5)$, and $z=1.19(1)$.
All exponents considerably differ from DP exponents.
To support these estimations, let us notice that the hyperscaling relation
$\frac{z}{2}=\Theta+\delta+\eta$ is satisfied by the above values.
It is rather surprising for us to observe that our values  of exponents 
$\beta$, $\Theta$ and $z$ are in a very good agreement with recent estimation
for the so-called bosonic version of a Pair Contact Process with Diffusion 
(PCPD)~\cite{KOCKEL}.
It would be interesting to examine whether this is only a numerical 
coincidence or there is a deeper relation between  these two problems.

\begin{figure}
\vspace{0cm}
\centerline{\epsfxsize=8cm
\epsfbox{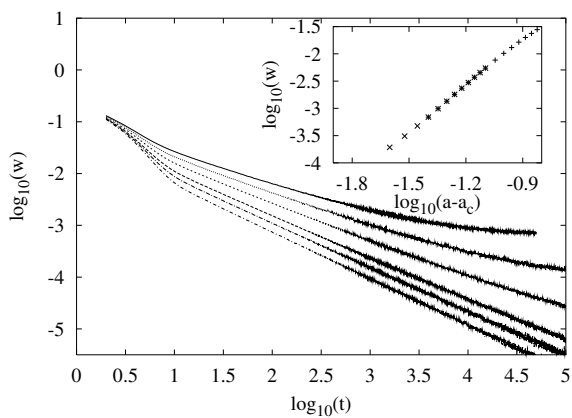}
}
\caption{
The time-dependent activity $w(t)$ as a function of time $t$ for
(from top) $a=0.11$, 0.09, 0.07 (critical point), 0.05, and 0.04
(Model II, $\gamma=1$).
The results are averaged over 200 independent runs and simulations were done 
for $L=5\cdot 10^4$.
Inset shows the logarithmic scaling of the the steady-state activity 
$w$ as a function  of $a-a_c$ with $a_c=0.07$ calculated for $L=2\cdot 
10^4 (+)$ and $L=5\cdot 10^4 (\times)$ (Model II, $\gamma=1$).
}
\label{timeadd}
\end{figure}

We also studied model II for fixed $\gamma$ and varying the amplitude $a$.
Monte Carlo simulations show that for $\gamma=2$ the critical behaviour of the 
model seems to be the same as for $\gamma=\gamma_c=1.74925$ and $a=1$.
It suggests that for a certain range of $\gamma$ the depinning transition 
belongs to the same universality class.
However, a different behaviour was observed for $\gamma=1$.
Indeed, in this case our simulations suggest (Fig.~\ref{timeadd}) that the 
depinning transition is characterized by the exponents $\beta\sim 3.0(3)$ and
$\Theta=0.62(5)$.
What is also interesting, the entire unbounded phase $a<a_c\sim 0.07$ 
seems to be 
critical with a power-law decaying order parameter $w\sim t^{-x}$, 
where $x\sim 0.8(5)$, and $x$ might slightly vary with $a$.
Such a novel behaviour is also worth further studies.

One of the future problems would be to check whether there are some other 
types of critical behaviour for this kind of SOS models.
Clearly, an important ingredient that determines the critical behaviour is the 
form of the function $p(h)$.
It would be interesting to explore some other functions (e.g., 
${\rm e}^{-\gamma h^2}$ or $[{\rm ln}(h)]^{-\gamma}$) for a possibly new 
behaviour.
Although we were not able to recover the BKPZ universality class within our
approach, there is still a possibility that for a certain choice of $p(h)$ such
a behaviour might appear.
It would be also interesting to check whether a new critical behaviour 
found in model II has a counterpart in
a synchronization transition in CML.

This work was partially supported by the Swiss National Science Foundation
and the project OFES 00-0578 "COSYC OF SENS".

\end {document}